\documentclass[11pt]{article}
\usepackage[utf8]{inputenc}
\usepackage{amsmath, amssymb}
\usepackage{graphicx}
\usepackage{hyperref}
\usepackage{geometry}
\usepackage{enumitem}
\usepackage{titlesec}
\usepackage{authblk,csquotes}
\usepackage{booktabs}
\hypersetup{colorlinks=true,allcolors=blue}

\usepackage{url}
\providecommand{\bibcommenthead}[1]{}      % 参考文献标题
\providecommand{\burl}[1]{}                % URL命令 - 输出为空
\providecommand{\url}[1]{}                 % \url命令 - 输出为空  
\providecommand{\doi}[1]{}                 % DOI链接 - 输出为空
           
\usepackage{setspace}
\setlist[itemize]{leftmargin=*, itemsep=0.5ex, topsep=0.5ex, parsep=0pt, partopsep=0pt}

\titleformat{\section}{\normalfont\Large\bfseries}{\thesection}{0.5em}{}
\titleformat{\subsection}{\normalfont\large\bfseries}{\thesubsection}{0.5em}{}
\titleformat{\subsubsection}{\normalfont\normalsize\bfseries}{\thesubsubsection}{0.5em}{}

\usepackage{xcolor}

\usepackage{soul}
\soulregister\cite7
\soulregister\ref7
\soulregister\onlinecite7

\usepackage{etoolbox}
\renewenvironment{abstract}{%
    \par\vspace{0.3em}%
    \begin{center}\textbf{\Large Abstract}\end{center}\vspace{0.1em}%
    \noindent
}{\par\vspace{0.5em}}

\usepackage{fancyhdr}
\title{\textbf{Research Paradigm of Materials Science Tetrahedra \\with Artificial Intelligence}}
\author[1,2]{\textbf{Shiyun Zhang}}
\author[2]{\textbf{Yibo Yao}}
\author[3]{\textbf{Haoquan Long}}%longhaoquan25@mails.ucas.ac.cn
\author[3]{\textbf{Dingwen Tao}}%taodingwen@ict.ac.cn
\author[3]{\textbf{Guangming Tan}}
\author[1]{\textbf{Wei-Hua Wang}} %whw@dimst.ac.cn
\author[1,2]{\textbf{Yuan-Chao Hu\thanks{Corresponding author: yuanchao.hu@dimst.ac.cn}}}

\affil[1]{Dongguan Institute of Materials Science and Technology, Dongguan, China}
\affil[2]{Songshan Lake Materials Laboratory, Dongguan, China}
\affil[3]{Institute of Computing Technology, Chinese Academy of Sciences, Beijing, China}
\date{\vspace{-1.5em}(Dated: \today)}

\begin{document}

\maketitle
% tighten space between title block (affiliations/date) and abstract
\vspace{-2.5em}

\begin{abstract}
The classical material tetrahedron that represents the Structure-Property-Processing-Performance-Characterization relationship is the most important research paradigm in materials science so far. It has served as a protocol to guide experiments, modeling, and theory to uncover hidden relationships between various aspects of a certain material. This substantially facilitates knowledge accumulation and material discovery with desired functionalities to realize versatile applications. In recent years, with the advent of artificial intelligence (AI) techniques, the attention of AI towards scientific research is soaring. The trials of implementing AI in various disciplines are endless, with great potential to revolutionize the research diagram. Despite the success in natural language processing and computer vision, how to effectively integrate AI with natural science is still a grand challenge, bearing in mind their fundamental differences. Inspired by these observations and limitations, we delve into the current research paradigm dictated by the classical material tetrahedron and propose two new paradigms to stimulate data-driven and AI-augmented research. One tetrahedron focuses on AI for materials science by considering the Matter-Data-Model-Potential-Agent diagram. The other demonstrates AI research by discussing Data-Architecture-Encoding-Optimization-Inference relationships. The crucial ingredients of these frameworks and their connections are discussed, which will likely motivate both scientific thinking refinement and technology advancement. Despite the widespread enthusiasm for chasing AI for science, we must analyze issues rationally to come up with well-defined, resolvable scientific problems in order to better master the power of AI.
\end{abstract}

\newpage
\begin{spacing}{1.15}

\section{Introduction}
%Introduce the problem, motivation, and background. Reference related work using \cite{key}.
Tracing back to the history of human civilization, critical materials always catalyzed societal revolution. In the Stone Age, natural resources like bone and wood were important for hunting and living. The emergence of metallic mixtures initiates the Bronze Age, which is followed by the Iron Age. After that, porcelain ceramics start to take the stage. By achieving higher processing temperatures given by using coal, steel smelting becomes feasible, and we enter the Industrial Era. The First Industrial Revolution kicked off the modern civilization and sped up the society evolution. By making the most of steel and associated products, the modern society was built worldwide. This lasts for about seven decades, taking over with the crucial usage of silicon, marking the beginning of the Information Age until nowadays. This streamlining highlights the crucial roles of developing and optimizing materials. What is the next breakthrough that is still waiting for next-generation material innovation~\cite{callister_materials_2018}.

\begin{figure}[t!]
\centering
\includegraphics[width=\textwidth]{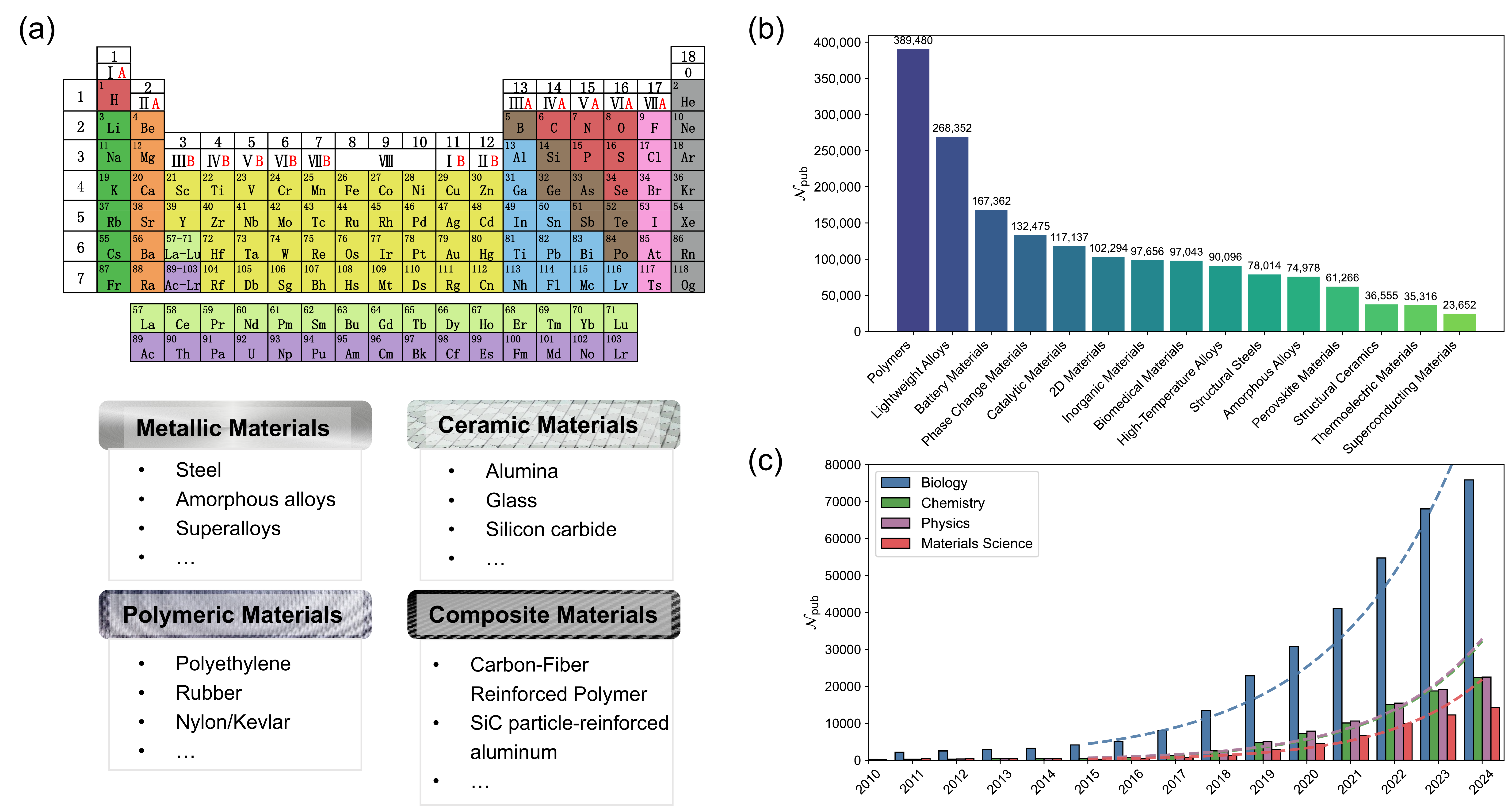}
\caption{
{\bf Overview of materials science research and AI integration}.
{\bf a}, The periodic table lays the foundation of various materials. The lower panel shows examples of materials of different types for illustration.
{\bf b}, The number of publications, $\mathcal{N}_{\rm pub}$, in a variety of materials science research fields.
{\bf c}, Yearly $\mathcal{N}_{\rm pub}$ for different subjects with additional search keywords "machine learning or artificial intelligence", demonstrating the research activities of integrating AI with these subjects.
}
\label{fig1}
\end{figure}

In principle, materials are carriers of science and technology. To realize versatile functionalities, a variety of materials have been developed or are undergoing development.
%for instance, metallic alloys, semiconductors, superconductors, polymers, inorganic compounds, energy materials, biological materials, to list a few. 
Materials science fundamentally categorizes materials based on their atomic composition and the primary types of chemical bonds governing their structure and properties, as exemplified in Fig.~\ref{fig1}a.
Intrinsically, all these materials are made of combinations of selected elements from the 118 chemical elements summarized in the periodic table (see Fig.~\ref{fig1}a). 
Regulated by the microscopic interatomic interactions, they show distinct macroscopic properties to enable broad applications and great potential.
In general, materials science and associated fields have two research routines. One is discovering new materials by trial-and-error with emerging fortune. The other is accumulating the fundamental knowledge by learning the underlying physical laws. The two pathways proceed with mutual promotion. For example, over the past century, the development of crystalline materials with structured theoretical frameworks and advanced microscopes has greatly enriched the natural science knowledge base~\cite{kittel2004introduction}.

To illustrate the scientific activities in materials science, we performed data mining over the literature database and depicted the number of publications ($\mathcal{N}_{\rm pub}$) on the topics of different materials in Fig.~\ref{fig1}b. Within this survey, it is not hard to see that generally there are more than $10,000$ papers published, which is linked to the paramount efforts from research teams all around the world. This number also roughly reveals the scale of the interested researchers in each domain.
Even though the constituents and compositions vary enormously, to the best of our knowledge, these fields are fairly dealing with three fundamental scientific questions:
\begin{itemize}
\item \noindent What are the physical principles ruling the accurate interactions among chemical elements?
\item Why can combinations of specific elements at certain composition can realize targeted functionalities?
\item How to efficiently navigate the material space to optimize matter design?
\end{itemize}

%%%%%%%%%%%%%%%%%%%%%%
%%%%%%%%%%%%%%%%%%%%%%

The past century has witnessed the combination of theory, modeling, and experiments in unveiling the mysteries beneath these questions with a specific material carrier. In particular, microstructure characterization techniques (Scanning Electron Microscope and Transmission Electron Microscope, etc.) and the spectrum measurements (X-Ray Diffraction, Raman Spectra, Neutron Scattering, etc.) grant unprecedented power to bridge the microscopic world with the macroscopic one. Looking back at the materials science literature, there is massive work addressing the structural features and related properties of different materials to gain new knowledge. This has evolved to be a research protocol in various fields, indicated, quite often, by the scientific writing strategy. The classical material tetrahedron was initiated in this streamline (see more discussion below).

% Because of the limited spatial-temporal resolution towards the microscopic world, computer simulations with either empirical or quantum-mechanical potentials have been playing a crucial role in materials science. 

Beyond experimental resolution limits, computer simulations play a crucial role in materials science by enabling access to atomic-scale mechanisms, extended time scales, and hypothetical configurations that are difficult or impossible to probe experimentally. Only in this way can we monitor the particle-level trajectories and the associated physical states. This brings fascinating scientific insights into understanding the underlying physical process otherwise overlooked. Opposite to the disappointments in some perspectives, the existing gaps among these aspects actually advance scientific research and technology innovation. The coexistence of experiment and modeling thus becomes meaningful, instead of replacement.
Meanwhile, computational science has been benefiting from the fast development of computational software and hardware. The algorithm evolution and the physical laws together boost natural science advancement. 

Nowadays, materials science stands at a critical juncture. For decades, this discipline has progressed through a symbiotic relationship between experimental observation, theoretical derivation, and computer simulation. This workflow has been effectively represented by the classical materials tetrahedron, which links processing, structure, properties, and performance. However, the design space for next-generation materials—ranging from complex alloys to molecular systems—is combinatorially vast, far exceeding the throughput of traditional trial-and-error methodologies.

In recent years, deep learning and artificial intelligence (AI) techniques brought outstanding breakthroughs in natural language processing and computer vision by integrating modalities~\cite{lecun_deep_2015,NIPS2017_attention,dosovitskiy2021imageworth16x16words}. It is gaining enormous attention in almost every field all around the world (see, for example, refs.~\cite{AI_science_2023,hao2026_AI_science}). In scientific research, its importance is also rocketing. The concept of ``AI for Science" is spreading widely and becoming the next competition hotspot. The most notable success of AI in research is the deployment of AlphaFold by the DeepMind team~\cite{Jumper2021HighlyAP}. Without language information processing, it integrates several components, including Transformer and deep graph learning, to capture the hidden patterns at the hierarchical scale to predict protein structure. This breakthrough brings confidence to other research fields, for example, materials science.

To gain insights from the implementation of machine learning or AI in scientific research, we also performed data mining in the literature database for different disciplines. As shown in Fig.~\ref{fig1}c, remarkably, the number of publications in Biology, Chemistry, Physics, and Materials Science all grow exponentially. There are considerably more trials of machine learning and AI in biology than the others, which may hint at the success of AlphaFold from the continuous efforts.
While materials science is still in its embryonic stage. This also indicates the unbounded potential of AI in future materials science research~\cite{cheng2026AIMaterials,Horton2025AcceleratedDD}.

Nevertheless, we must bear in mind the fundamental differences between materials science and other fields like natural language and biology. 
Materials systems often involve multi-scale interactions, sparse datasets, and strict physical constraints (e.g., symmetry and conservation laws) that general-purpose AI models are absent from.
The effective token design and block-wise processing are very challenging in materials science. The application of SMILE (Simplified Molecular-Input Line-Entry System) is a typical string notation of molecular systems for chemical language models~\cite{Skinnider2024SMILES}.
Another crucial feature associated with materials science is the data scarcity with enriched physics~\cite{PINN2021NRP}. Balancing the trade-off between the amount of data and the underlying physics is critical, which shall rely on the proposal of efficient data sampling strategies. This is the most prevalent in computational studies of rare-event phase transformation kinetics by enhanced sampling~\cite{Invernizzi2020UnifiedAT}.
As data accumulation is quite expensive and time-consuming in materials science, nowadays, it is impossible to quickly gain enough high-quality data for AI implementation. The smart synthesis automation system is capable of agitating this process~\cite{Szymanski2023AutonomousLA}, subject to the designed sampling strategy. Being faced with the fast growth of AI for science, it may be quite beneficial to come up with an effective research paradigm to search for suitable pathways before diving deeply into the research field.

To harness the full potential of AI in materials science, it would be quite beneficial to refine our research paradigms. Merely applying existing computer science algorithms to material datasets is insufficient.
In this work, we systematically investigate the current research paradigm represented by the classical material tetrahedron that demonstrates the structure-property-processing-performance-characterization relationship. By carefully considering the research philosophy, we try to propose a new tetrahedron for both AI for materials science and the AI research itself. By discussing the tetrahedral ingredients, we hope to unveil possible research targets and strategies to motivate future integration of AI techniques with materials science challenges.
In addition, we also initiate our research idea of material network science to solve materials science problems with limited data and to unravel the hidden patterns in higher-dimensional latent spaces. Once the resolvable scientific problem is well defined, these paradigms shall provide effective routes to make progress.

To clarify the conceptual relationship between these paradigms, we summarize the three tetrahedral frameworks discussed in this work in Table~1, highlighting their corresponding components across classical materials science, AI-driven materials research and engineering, and AI methodology itself.

\begin{table}[t]
\centering
\caption{Conceptual correspondence among the three tetrahedral research paradigms discussed in this work. (MSE: Materials Science \& Engineering)}
\begin{tabular}{lcc}
\hline
\color{black}
\textbf{Classical MSE} & \textbf{AI for Materials} & \textbf{AI Research} \\
\midrule
Structure        & Data      & Architecture \\
Processing       & Model     & Encoding \\
Properties       & Potential & Optimization \\
Performance      & Agent     & Inference \\
\midrule
\textbf{Characterization} & \textbf{Matter} & \textbf{Data} \\
\bottomrule
\end{tabular}
\end{table}

\section{Results and Discussion}
%Show your results, tables, and figures. Use \ref{fig:example} to reference figures.

\subsection{The classical material tetrahedron}

\begin{figure}[t!]
\centering
\includegraphics[width=\textwidth]{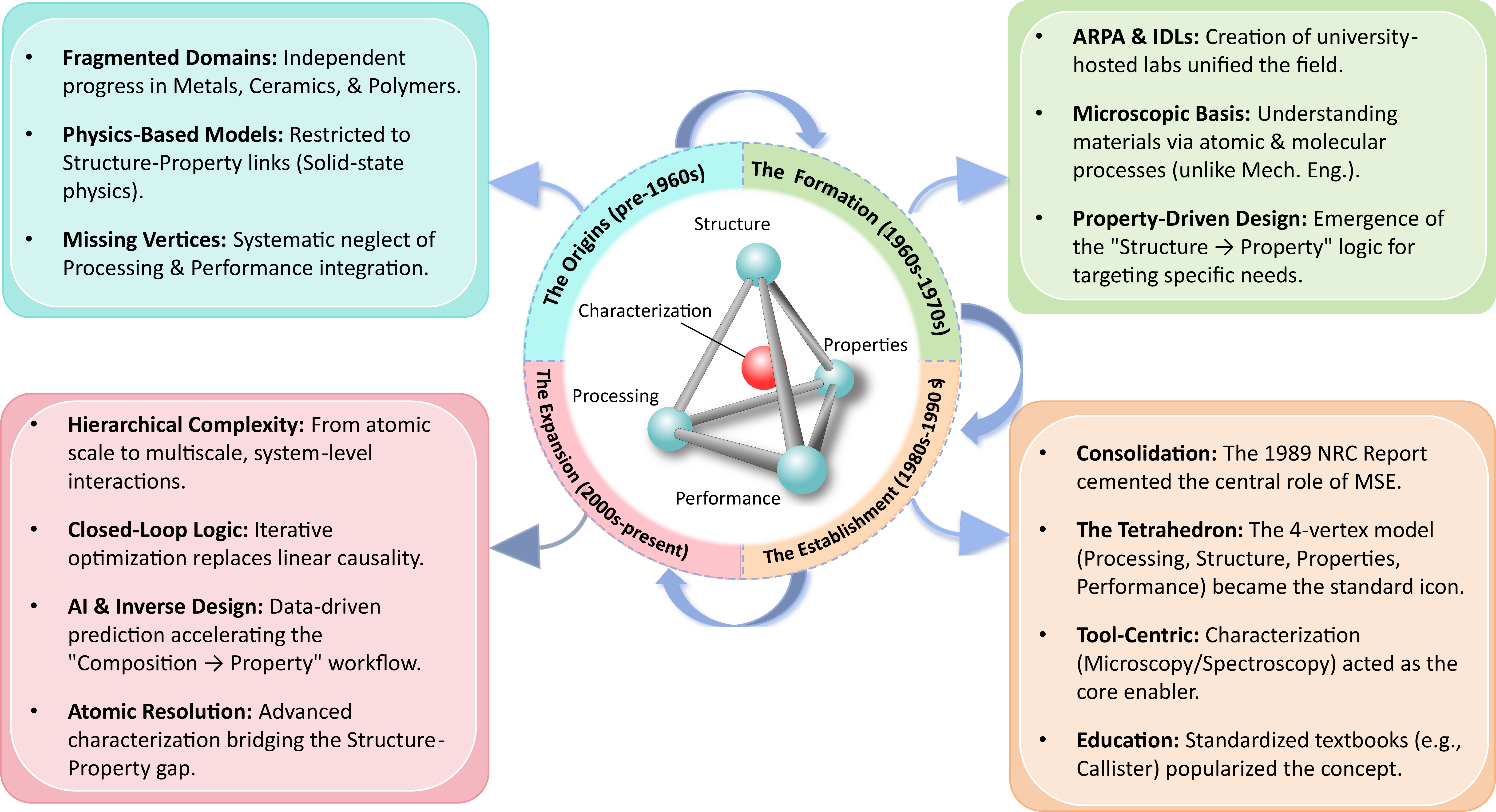}
\caption{
{\bf The classical material tetrahedron}.
The historical overview of the research diagram of the structure-property-processing-performance-characterization relationship. There are roughly four stages with the demonstrated key components in summary.
}
\label{fig2}
\end{figure}

In materials science research, identifying the key problems to address is of crucial significance, together with the suitable protocol. Nevertheless, the research goal is pretty straightforward: designing applicable materials. With the accruing physical knowledge and experimental apparatus, there appears a general, while useful research paradigm, dictating the structure-property-processing-performance-characterization relationship in a tetrahedral unit~\cite{NRC_MSE1990s_1989,NRC_MaterialsAndMansNeeds_1974,Meyers_Chawla_2025}. This classical material tetrahedron is shown in Fig.~\ref{fig2} with some key historical information accompanied. This empirical summary somehow demonstrates the knowledge-gaining process of materials science.
Generally speaking, the scientific research can focus on one of these aspects or any combination of these components. We are now discussing the evolving four stages from a historical view.

\subsubsection{Historical Foundations: Pre-1960s Fragmentation}
Prior to the 1960s, materials research exhibited significant disciplinary fragmentation. 
Contemporary materials science departments predominantly evolved from specialized metallurgy or ceramics engineering departments, reflecting the historical emphasis on these material classes during the 19th century and early 20th century. 
While fundamental physical and chemical principles govern metals, ceramics, and the emerging field of polymers, research within these domains progressed largely independently, establishing distinct intellectual traditions. 
Theoretical modeling during this era focused predominantly on structure-property relationships, exemplified by foundational work in lattice theory and solid-state physics. Another crucial catalyst is the fast development of the microscope techniques.
This paradigm, while crucial for understanding intrinsic material behavior, systematically neglected the integration of processing parameters and explicit performance requirements---key elements later formalized in the material tetrahedron framework.

\subsubsection{Disciplinary Formation: Codification of the Tetrahedron Paradigm (1960s--1970s).}
The consolidation of materials science as a unified discipline gained significant momentum during the 1960s, driven by institutional initiatives in the United States. 
The Advanced Research Projects Agency (ARPA) catalyzed this transformation through strategic funding of university-hosted laboratories in the early 1960s, explicitly aiming ``to expand the national program of basic research and training in the materials sciences"~\cite{Name_change_2015}. This institutional support fostered interdisciplinary collaboration across traditional material domains. 
A pivotal conceptual advancement emerged in the National Materials Advisory Board (NMAB) report, which formalized ``materials science \& engineering" (MSE) as an integrated field centered on the fundamental composition-structure-properties relationship~\cite{NRC_MSE1990s_1989}. 
By the early 1970s, this conceptual framework underwent critical expansion through the systematic incorporation of processing and performance parameters. 
This evolution culminated in the geometric representation of materials research as an interconnected tetrahedron---a paradigm capturing the quintessential interdependence of Processing, Structure, Properties, and Performance.

%%%%%%%%%%%%%%%%%%%%%%
%%%%%%%%%%%%%%%%%%%%%%

\subsubsection{Paradigm Establishment: Canonical Representation and Institutional Adoption (1980s--1990s).}
The 1980s witnessed the formal codification of the tetrahedron paradigm within academic discourse, transitioning from a conceptual framework to a canonical representation. Standardized textbooks systematically defined the four interdependent pillars—composition and structure, synthesis and processing, properties, and performance—thereby establishing the foundation for modern materials science pedagogy. This formalization marked a fundamental shift from empirical trial-and-error approaches toward rational design methodologies. Within this framework, structure deterministically governs properties, synthesis and processing control the structure, and performance requirements drive iterative optimization cycles. The geometric formalism of the tetrahedron explicitly captured these causal relationships, enabling predictive modeling of material behavior. Institutional recognition culminated with the 1990 U.S. National Research Council (NRC) report~\cite{NRC_MSE1990s_1989}, Materials Science and Engineering for the 1990s, which enshrined this framework as the fundamental paradigm of the field. The NRC explicitly emphasized the role of the tetrahedron in integrating synthesis and processing, structure and composition, properties, and performance as coequal elements of materials research.

\subsubsection{Refinement and Expansion: Multiscale Integration and Digital Transformation (2000--Present)}
The 21st century has witnessed profound refinement and expansion of the materials tetrahedron paradigm, driven by conceptual advances and technological breakthroughs. Conceptually, the framework underwent three fundamental shifts. 
Firstly, the focus expanded from atomic-scale structure to hierarchical microstructures spanning multiple length scales, acknowledging that material behavior emerges from complex multiscale interactions. 
Secondly, the emphasis shifted from macroscopic properties to underlying multiscale mechanisms, prioritizing mechanistic understanding over phenomenological observation. 
Thirdly, the ultimate criterion evolved from component-level metrics to system-level performance, integrating service environments, reliability analysis, and failure mode prediction. These developments transformed the paradigm from linear causality to closed-loop interdependence, where feedback between elements drives iterative optimization.

Technologically, two revolutions have further reshaped the tetrahedron. Advances in characterization, such as aberration-corrected transmission electron microscopy and atom-probe tomography, enabled atomic-scale resolution of structure-property linkages. Simultaneously, the computational revolution, catalyzed by the Materials Genome Initiative (MGI) in 2011~\cite{MGI_2011, MGI_2019}, digitized the tetrahedron. This transition facilitated a shift from description to prediction and optimization, moving the field from simple structure-property correlations to inverse design paradigms. The integration of AI further accelerates the prediction of composition, processing, and property relationships, marking the beginning of a data-driven era in materials research~\cite{data_driven_MS_2019, ML_MS_2019, small_data_2023}.

\subsection{AI-augmented materials science}
%%%%%%%%%%%%%%%%%%%%%%
\begin{figure}[t!]
\centering
\includegraphics[width=\textwidth]{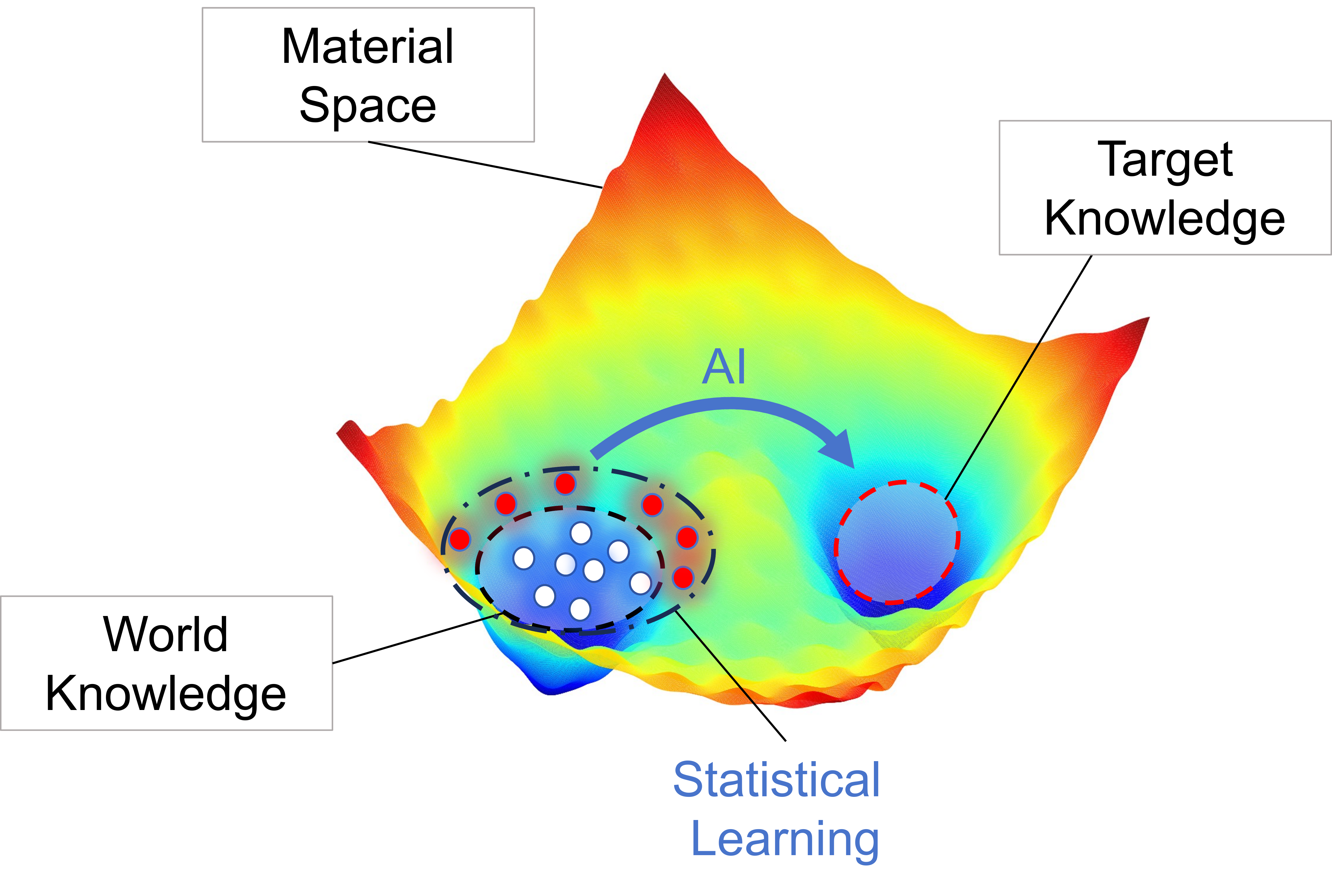}
\caption{
{\bf Schematic learning landscape of AI for materials science}.
The known materials and the associated knowledge (World Knowledge) are conceptually shown within a single local minimum for simplicity. It is only an iceberg in the potential material space. Statistical learning usually figures out the hidden pattern locally, bringing limited power to expand the knowledge boundary. Target knowledge is separated into another local minimum to show the desire of grand breakthrough. AI is promising to guide navigation on the landscape with two research targets: domain exploration and pathway optimization.
}
\label{fig3}
\end{figure}
%%%%%%%%%%%%%%%%%%%%%%

Learning from data to gain experience is by no means a new topic, but the way to do it can be. 
%Looking into the traditional research strategy, we are always accumulating data from either experimental measurements or computer simulations or theoretical calculations or their combinations and then analyze them with the learned domain knowledge to acquire new things. 
Traditionally, research strategies have relied on the accumulation of data through experimental measurements, computer simulations, or theoretical calculations, followed by analysis based on established domain knowledge.
This process relies greatly on the way of data acquisition, from which physical laws can be derived. 
Nowadays, we are turning into another phase: data-centric. That is, can critical insights be revealed directly from sophisticated data learning? This philosophy is intrinsically different from the traditional one. It shall crack the border between fields and promote interdisciplinary collaborations, for example, data science and materials science. AI is the pioneering tool to take full advantage of these data, demonstrating data-driven and AI-augmented materials science~\cite{cheng2026AIMaterials, Horton2025AcceleratedDD}. 
As this topic is attracting immense research attention, there are too many specific ways to list here. However, regarding the three critical questions introduced above, there exist fundamental thoughts behind the technology evolution. Although AI is very powerful in many cases, it is absolutely not an alchemist. It creates intelligent ways for information storage, organization, and retrieval. To better integrate AI with materials science, it would be beneficial to hold some enthusiasm and deeply think about the fundamental philosophy.

The schematic illustrated in Fig.~\ref{fig3} conveys our understanding of the fundamentals of AI-reinforced materials science. On the conceptual level, we summarize all the known knowledge from all accumulated data and locate it in a local minimum area ("World Knowledge"). This convergence is not realistic, but demonstrates the waiting for the next-breakthrough from what we currently know in each field. It is fair to argue that our current knowledge about natural science is still rather limited than what Nature grants us. With suitable labels for the existing data, in whatever format, a mapping function is intuitively expected to connect them (e.g., ref.~\cite{hart_machine_2021}). With simple mathematical equations, we can handle it directly. To capture more complicated relationships, statistical learning comes into play.
%\hl{statistical learning includes un-supervised machine learning. For example, PCA and t-SNE.}
Based on statistical theories with sufficient data, various models, such as linear regression, logistic regression, support vector machine, decision tree, and some boosting and ensemble methods, can be developed to establish the hidden connection between input data and the associated labels. With a confident model training or data learning, the resultant models are used for different downstream tasks. If only the gained extrapolations were considered, the knowledge border could be pushed forward by statistical learning. Usually, this generalization capability can be limited.
More significantly, the potential material space can be much larger than currently expected. There are more precious regions waiting to be explored.

%These observations and limitations motivate the development of deep learning through designing neural networks. Unlike statistical machine learning, deep learning encapsulates internal powerful feature transformation and optimization with multiple layers to capture the hidden linear and nonlinear couplings. Together with Transformer, they serve as the critical components of the current AI techniques.
These limitations motivate the adoption of deep learning~\cite{lecun_deep_2015}. Unlike traditional statistical methods, deep learning employs multi-layer neural networks to perform automatic feature extraction and transformation, capturing complex linear and nonlinear couplings within high-dimensional data. Architectures such as Transformers have further revolutionized this capability, serving as critical components in modern AI techniques.

The major driving force for us to develop AI for science is that we believe that AI will finally bring something new. That is, we have a belief that there exists at least an unknown domain that carries what we desire, naming the "Target Knowledge" in Fig.~\ref{fig3}. Thus, the first research target of AI for science is not just ``exploitation" of known data but also ``domain exploration": to confirm the existence possibility and identify its location on the landscape. This is highly aligned with well-defined, resolvable problems by implementing AI, i.e., ``AI-ready" problems.
Once certified, the AI task becomes ``pathway optimization": to figure out the best pathway with successive minimal local barriers from the current knowledge to the target domain. One typical example is the discovery of new materials.
Physically, these tasks are in some sense similar to the phase transformation of materials, in which a phase will transform to another along the minimal free energy barrier pathway driven by thermodynamics and kinetics.
This design will help clarify the applicability of AI in different materials science scenarios.

\subsection{Tetrahedron of AI for materials science}

%%%%%%%%%%%%%%%%%%%%%%
\begin{figure}[t!]
\centering
\includegraphics[width=\textwidth]{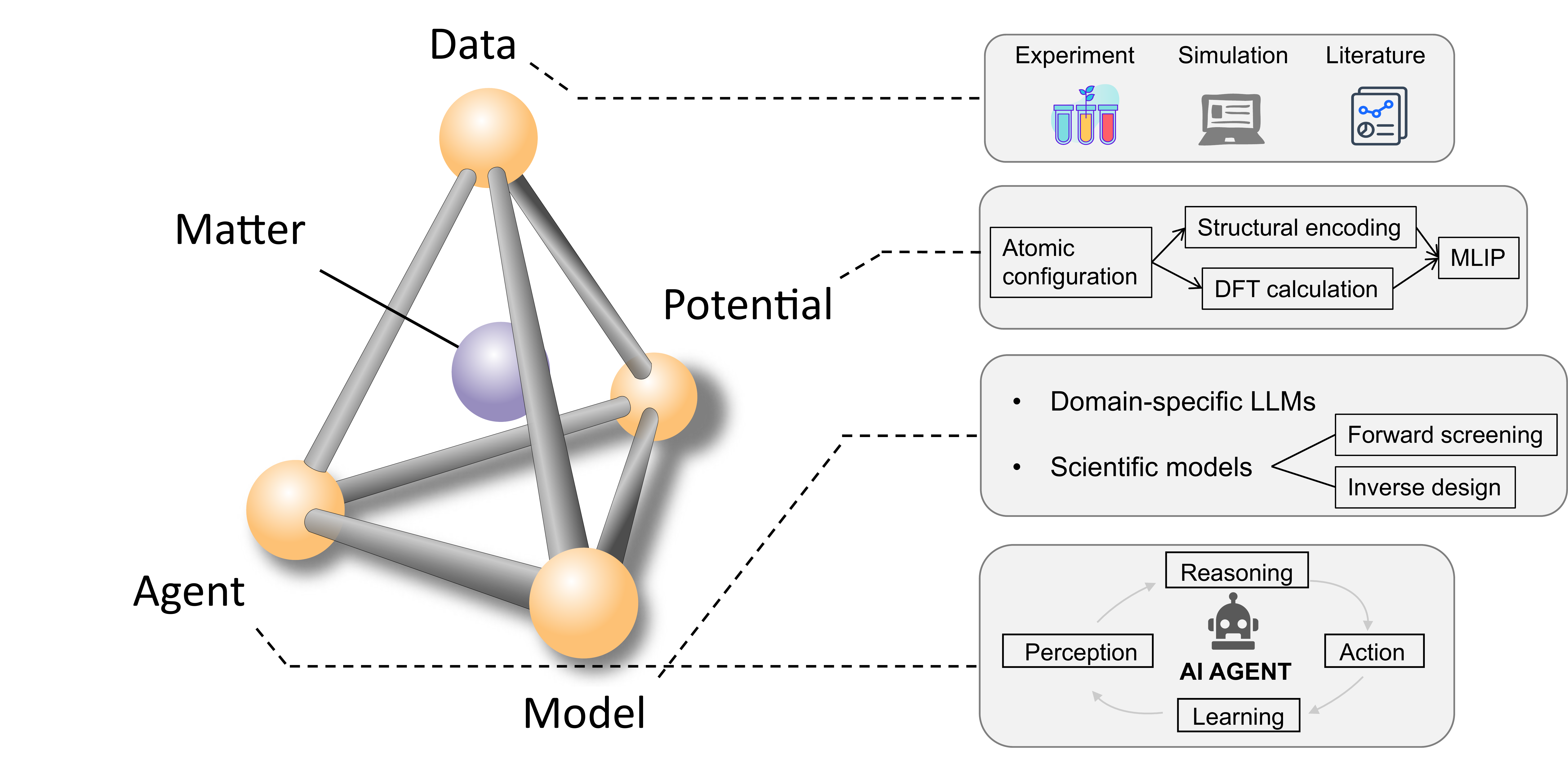}
\caption{
{\bf AI-augmented material research tetrahedron}.
With Matter centered, the tetrahedral network has four components, i.e., Data, Model, Potential, and Agent. These vertices and their combinations are all significant for AI-augmented materials science. Some extended components are included for illustration.
}
\label{fig4}
\end{figure}
%%%%%%%%%%%%%%%%%%%%%%

To stimulate optimized research strategies of implementing AI for materials science, we shall rationalize the key fundamental components in pursuing it. Clarifying these will help make solid research strategies to resolve the critical issues. After literature review and practice engineering, we propose the four components associated with the centric Matter: Data, Model, Agent, and Potential. Their relationship is organized in a tetrahedron in Fig.~\ref{fig4}. In the following, we are going to provide a high-level discussion on the details and their connections.

The center is Matter, which includes all aspects of scientific insights and problems associated with all different kinds of materials (see some examples in Fig.~\ref{fig1}a).
One typical example is amorphous alloys or called metallic glasses~\cite{wang_metallic_2025,zhang_constructing_2025}. The associated scientific problems can be the glass-forming ability and soft-magnetic properties, and so on.
The goal is to invent new materials and understand the underlying physics. The other four components and their combinations are of crucial significance in achieving it. In turn, what these components can do actually determines the effective AI-ready materials science problems.

We start from Data, which can be treated as the fuel of modern AI. In principle, both the quantity and quality of data are critical. 
On one hand, when the amount of data is small, there exists deep physics we need to learn from this limitation~\cite{PINN2021NRP}. As the space could be quite rugged, the resultant gradient in the latent space is highly likely to change sharply. It requires expertise to accomplish the tasks; human brain learning is essential --- only a super-big model can deal with this situation. This dictates the situations in natural science research even nowadays. From the perspective of AI, this is also reasonable, as the human brain has way more parameters and complex structures than the most state-of-the-art AI models.
On the other hand, when the data amount is massive, there are chances of emerging patterns directly from the data in the proper latent space. A typical example is the modern large language models (LLMs)~\cite{xiao_densing_2025}. There is not necessarily rich physics or rules in the corpus.
Therefore, there is a hidden trade-off relation between data amount and model complexity. We should balance the data quality and quantity in materials science. From this respect, the sampling strategy is of tremendous importance. We discuss this from the Hoeffding Inequality:
In regimes of data scarcity, which are common in experimental science, the latent space is often rugged with discontinuous gradients, necessitating the incorporation of deep physical priors or expert knowledge to guide learning. The success rate is, in fact, very low. Conversely, in regimes of massive data abundance, emerging patterns can be extracted directly from the data distribution without explicit rule-based programming. To rigorously quantify the relationship between sample size and generalization capability, we invoke the Hoeffding Inequality:
\begin{equation}
P(|E_{\rm in} - E_{\rm out}| > \epsilon) \leq 2e^{-2\epsilon^2N},
\label{eq1}
\end{equation}
where $E_{\rm in}$ and $E_{\rm out}$ are the in-group and out-of-group error, respectively. $\epsilon$ is the error tolerance and $N$ is the number of in-group data points. It claims that increasing $N$ should effectively reduce the learning error so that knowledge is able to be learned from the unseen data. Thus, how to increase $N$ is critical, reverting to the importance of the sampling strategy. This is strongly related to the current understanding of the topic under investigation, where human creativity still matters critically. A typical example is enhanced sampling in studying the phase transformation of materials~\cite{Invernizzi2020UnifiedAT}.
In addition, there are many technical points to be addressed or to be aware of~\cite{wang_metallic_2025}, such as data standardization, data recording, data cleaning, data extraction, data storage, data representation, and so on. These practical problems are very important to consider, while at a high-level, here we also emphasize the importance of physics-driven data science. This will reveal the hidden patterns underneath, which will reduce the computation and model complexities.
Not only serving as foundations for model and application development, but data also provides efficient ways to gain insights.

The second vertex is Model, which represents the computational architectures designed to navigate the vast material space efficiently. The primary objective of the Model vertex is to reduce the dimensionality of the search space, thereby accelerating the discovery of materials with targeted functionalities. 
Generally, there exist two types of such models. The first one is the domain-specific LLMs. They absorb massive tokens from the scientific corpus and provide next-token generation during inference. These models aim to generate necessary knowledge information from scientific literature. This direction is straightforward with the development of LLMs. The second is the scientific (large) models.
Methodologically, these models can be broadly categorized into two paradigms: forward screening and inverse design. Forward screening models function as high-throughput surrogates, taking a material structure as input and predicting its properties (e.g., band-gap, formation energy)~\cite{mechant_scaling_2023}. This discriminative approach allows researchers to rapidly filter candidate libraries. Conversely, inverse design employs generative models—such as variational autoencoders (VAEs), generative adversarial networks (GANs), and diffusion models—to map desired properties directly to new chemical structures. A recent example is MatterGen~\cite{zeni_generative_2025}. By learning the distribution of stable materials, these models can generate novel candidates that do not exist in training databases, effectively solving the ``inverse problem" of materials discovery.
Conceptually, a model functions as a high-dimensional approximator in the latent space, transforming feature matrices into target predictions. The performance of the model is critically dependent on the inductive biases encoded in its architecture and the expressiveness of the feature engineering.

The next one is Potential, specifically referring to Machine Learning Interatomic Potentials (MLIPs) or equivalent terminologies. 
Here, we single out MLIPs with the aim of emphasizing their importance in bridging multi-scale computer simulations to power the development of the above Model.
In materials science, computer simulations are playing crucial role in capturing physical insights at the atomic scale. It helps fundamental knowledge acquisition.
There are many simulation methods spanning over a large range of time scales and length scales. Here, we focus primarily on density-functional theory (DFT) calculations and molecular dynamics (MD) simulations. While DFT is more accurate in describing the inter-particle interactions for various systems, it suffers from a rather limited number of particles and physical time to be simulated. Meanwhile, MD is capable of simulating much larger systems in microseconds, especially with GPU-acceleration, but relies seriously on the inaccurate empirical potentials. Previously, researchers developed these potentials by relying on DFT calculations with calibrations from experimental characterizations, but it is only feasible for simple systems. It is more powerful to simulate analytical interactions, such as the Lennard-Jones pair potential. Nowadays, the collaborations of DFT, deep learning, and MD generate new possibilities to model more complicated systems that are relevant to real applications, for example, liquid/solid electrolytes and multi-component alloys~\cite{mace_off_2025,batzner_e3nn_2022,Liu2024TheAS,xie_GNN_2018}. The underlying many-body interactions can be captured. MLIP effectively bridges DFT and MD, which greatly extends our capabilities to carry out more powerful simulations to gain critical insights. The universal foundation MLIP is still under development.
The development of modern MLIPs is currently defined by three critical technical dichotomies. The first is the choice of representation: local environment descriptors versus graph descriptions. Local descriptors rely on fixed-cutoff atomic environments and are computationally efficient, whereas graph neural networks utilize message-passing mechanisms to capture complex, many-body interactions, often yielding higher accuracy at a higher computational cost. The second is the treatment of symmetry: invariance versus equivariance. While energy is a scalar invariant under rotation, forces are vector quantities that must rotate with the system. State-of-the-art architectures now inherently enforce $\text{E}(3)$-equivariance, ensuring physical consistency without data augmentation. The third dichotomy is the scope of applicability: universal versus custom. Universal foundation models (e.g., M3GNet, CHGNet, MACE) are pre-trained on massive databases like the Materials Project to provide "out-of-the-box" simulations for nearly any element combination, whereas custom potentials are trained on small, domain-specific datasets via active learning to achieve maximum precision for a specific chemical system.
By singling out Potential as a distinct vertex, we emphasize the unique role of dynamics and time-evolution in understanding material behavior, which complements the static predictions of the Model vertex.
Benefitting from Data, MLIP will help fundamental knowledge acquisition, which in turn will help model design. The knowledge-graining process will play a more critical role in future competitions between AI and human beings.

The last component is Agent. It is attracting considerable attention empowered by AI~\cite{durante2024agentAI}. With various LLMs and functional tools, versatile agents have been developed for different modules of a task. Usually, tools or functions for executable tasks are encapsulated in an agent. 
By modularizing a big task into efficient small tasks, multiple agents can work collaboratively to accomplish the task. This idea has been widely used in literature mining, experimental design, and simulation prediction.
Automating the workflow with well-designed agents will be promising for next-generation research protocol (see, for example, ref.~\cite{ghafarollahi2025automating}). In principle, all the tasks related to data, model, and potential can somehow be realized by agent design, whether dominating or assisting.
This, meanwhile, demonstrates its critical reliance on the backend tools. Although with the criticalness of agentic automation workflows, developing the models should be paid more attention to in research.
The Agent vertex signifies a shift from passive computational tools to active research assistants that can navigate the scientific discovery process.
In the near future, a first principle thought would be replacing any steps with human intervention by an executable agent. The involvement of researchers will be plummeting, while the opposite will be true for versatile agents. Deeper thinking, optimized scheduling, and sophisticated analysis will be key tasks of researchers or operators.

Just like the classic material tetrahedron, the edges of the material tetrahedron of AI for materials science also represent pairwise interactions, such as the Data-Model loop (active learning) or the Potential-Agent interface (autonomous simulation). Higher-order combinations, such as the integration of Data, Model, and Agent, form the basis of self-driving laboratories. This structured decomposition helps identify missing links in current research workflows and provides a clear schematic for designing comprehensive AI-for-Science strategies. In fact, how to efficiently combine them will be key to this success.

\subsection{Tetrahedron of AI research}

Rome was not built in a day. This definitely applies as well to the development of AI. Even though the present AI is very powerful, especially the multi-modal LLMs, there are tremendous research efforts that have been devoted to them. From simple statistical learning models like logistic regression to complex deep neural networks, there has been a quick pace in AI development over the past 40 years. Here, we are not going to extend this history discussion, but emphasize the importance of AI research aligned with the materials science domain. 
As discussed above, there exist fundamental differences between LLMs and materials science research. We shall reconsider the direct applicability of modern AI techniques in specific materials research. How AI can be adapted to materials science is of crucial importance. For example, the autoregressive nature of next-token generation is quite successful in the language and image domains; can it be shifted to materials science? What is the fundamental logic difference that should be captured so that domain-specific AI can be developed? What are the main physical rules that will guide AI development for materials science? Besides the hybrid ``AI + Materials" research, AI research is still strongly desired by the materials science community. In the following, we propose a new research paradigm in the format of tetrahedron for such AI research.

The center of the research tetrahedron is Data, which has been extensively discussed above. 
There are two crucial research targets: (i) learning from data to get the underlying physical rules that can be embedded in the AI model, which is also of paramount importance to compensate for the lack of data; (ii) figuring out the key differences and similarities between the domain-specific data and the tokenizable texts and images. At this moment, proper strategies of data engineering, data science, and data generation should attract immense research attention to lay a solid foundation for model development.

%%%%%%%%%%%%%%%%%%%%%%
\begin{figure}[t!]
\centering
\includegraphics[width=\textwidth]{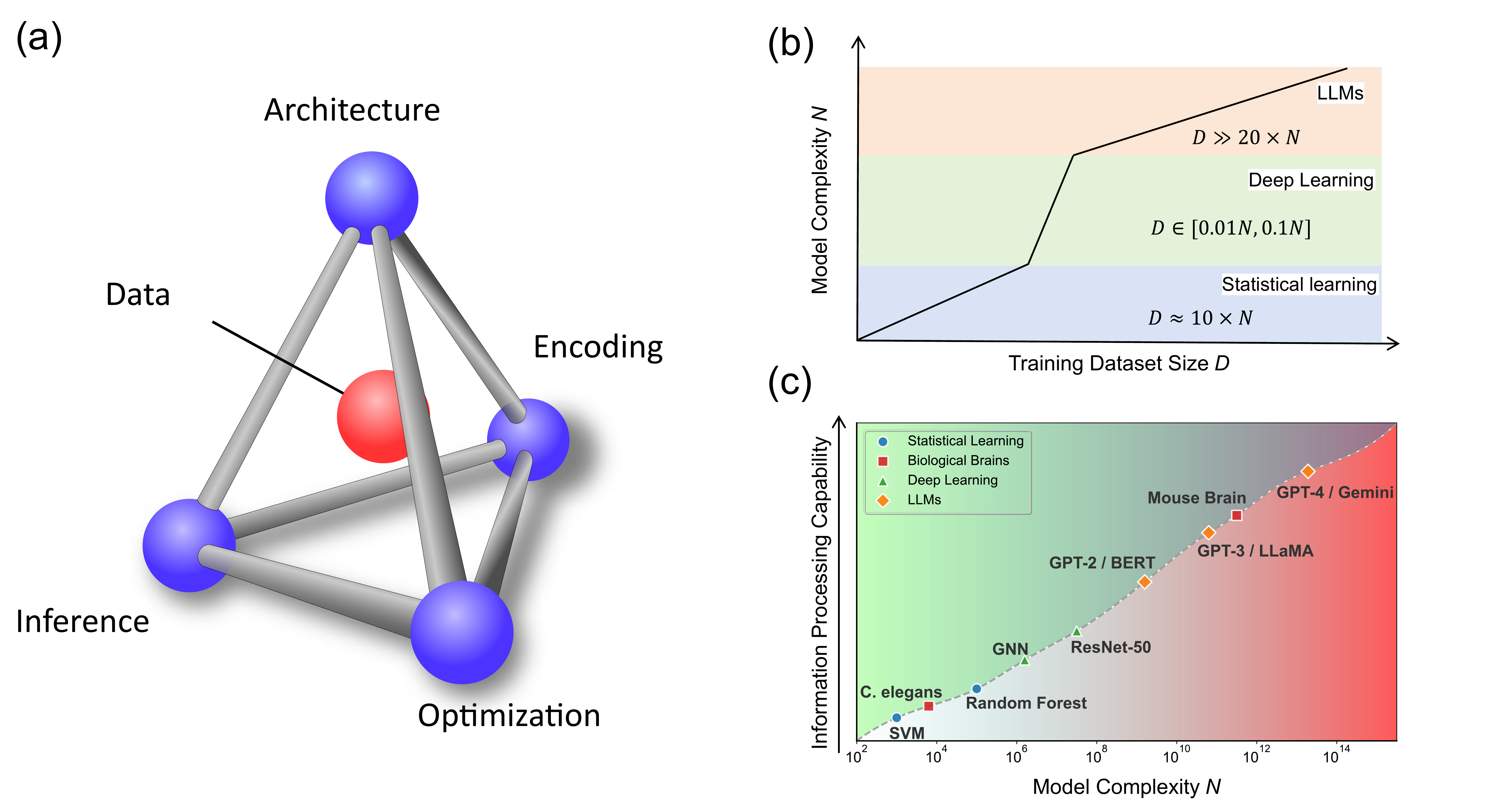}
\caption{
{\bf Tetrahedron for AI research}.
{\bf a}, There are four vertices centered at Data to make up the tetrahedron, i.e., Architecture, Encoding, Inference, and Optimization.
{\bf b}, Schematic relationship between model complexity and the sufficient size of the training dataset. Three regimes are shown: statistical learning, deep learning, and LLMs.
{\bf c}, An illustration depicting the capability of a model in processing input information. With increasing model complexity, it can process information in a wider range of complexity, with the upper bound extended.
Advanced LLMs like GPT-5/Gemini can process both simple and difficult requests, while statistical learning like SVM has limited capability with specific requirements.
}
\label{fig5}
\end{figure}
%%%%%%%%%%%%%%%%%%%%%%

We extend the discussion on Data to concern its valid necessity for model development. This is tightly linked to model generalization, and the growing enthusiasm on large foundation models.
The relationship between the sufficient training dataset size $D$ (measured in samples or tokens) and the number of trainable parameters $N$ has undergone significant paradigm shifts across different eras of machine learning. We categorize these relationships into three distinct regimes based on the governing theoretical frameworks, as illustrated in Fig.~\ref{fig5}b.
Firstly, in the classical regime (e.g., support vector machines, logistic regression), the relationship is governed by the Vapnik-Chervonenkis (VC) theory~\cite{Vapnik1998}. To guarantee a bounded generalization error $\epsilon$ with probability $(1-\delta)$, the sample complexity $D$ scales linearly with the VC-dimension $d_{\rm VC}$ (which is approximately proportional to $N$ for many architectures): 
\begin{equation}
D \propto \frac{d_{\rm VC} + \ln(1/\delta)}{\epsilon^2} \approx \mathcal{O}(N)
\end{equation}
Empirically, this regime adheres to the rule of thumb where $D \gg N$ (typically $D \approx 10N$) to avoid the curse of dimensionality and overfitting.
Secondly, modern deep neural networks (e.g., ResNets, VGGNet) operate in the interpolation regime, characterized by the double descent phenomenon~\cite{DGD_2019_pnas}. Here, models perform well even when $N \gg D$ due to the implicit regularization induced by Stochastic Gradient Descent (SGD) and optimized manifold-like parameter space because of large $N$.
In this regime, strict proportionality is abandoned. For instance, ResNet models trained on ImageNet-1K (1.28 million images) range from 11.7 million parameters to 60.2 million parameters, yielding $N \in [10D, 100D]$~\cite{resnet2016}. It leverages the inductive bias of convolutions to generalize despite the massive capacity.
Lastly, for Transformer-based LLMs, the relationship is formalized by the Scaling Laws. Let $\mathcal{L}(N, D)$ denote the cross-entropy loss. The dependence follows a power-law distribution~\cite{kaplan2020scalinglawsneurallanguage,hoffmann2022trainingcomputeoptimallargelanguage}:
\begin{equation}
	\mathcal{L}(N, D) = E + \frac{A}{N^\alpha} + \frac{B}{D^\beta},
\end{equation}
where $E$ is the irreducible entropy of natural language, and $A, B, \alpha, \beta$ are constants. Under a fixed compute budget $C \approx 6ND$ (in FLOPs), the Chinchilla optimality condition requires balancing the gradient contributions from model size and data size~\cite{hoffmann2022trainingcomputeoptimallargelanguage}:
\begin{equation}
	-\frac{\partial \mathcal{L}}{\partial N} \approx -\frac{\partial \mathcal{L}}{\partial D} \implies D^\star \approx 20 N^\star.
\end{equation}
This implies a linear optimal scaling where data tokens should be approximately 20 times the parameter count. However, recent trends (e.g., LLaMa3) pursue Inference-Optimality, pushing $D \gg 20N$ to minimize inference latency rather than training cost~\cite{grattafiori2024llama3herdmodels}.
Advancing model architecture is an effective way to better take advantage of the data resource for learnable problems.

Here we explore further on the information processing capability of models. Such information complexity is dependent on the learned messages from the input data. We show an illustration in Fig.~\ref{fig5}c. 
State-of-the-art models with different complexities should be able to process information at different levels of complexity. For instance, LLMs like GPT-5 and Gemini with trillion-level parameters ~\cite{luo2025evaluating} are capable of processing both simple and difficult tasks. They learned rich patterns from the large input dataset. Enormous amount of information combinations can be retrieved by tuning the versatile input prompts. They can even be more powerful than some biological brains, like mouse brain with $\sim 10^{12}$ synapse. 
Nevertheless, statistical learning models like Support Vector Machine (SVM) and Random Forest can process comparatively limited information under necessary conditions, for example, obeying certain probability distribution or meeting some boundary conditions. 
More importantly, in materials science research, there is very few data available, which requires super intelligence to process the information. This, to some degree, falls into the capability of human brains. This can be understood from the current routine of scientific research as discussed above. However, it can be hard for other models, like statistical learning and LLMs. This demonstrates the lower-than-expected power of these models in dealing with hard materials science problems under the condition of insufficient data. Another message to convey is the still-limited power of LLMs than human brains considering the number of neurons, the way of pre-training, and the learning architecture. The data-physics balance should be carefully considered and on such a basis more sophisticated models are waiting to be developed.

The first vertex we are going to discuss is Architecture. From our understanding, the progress in AI is quite often driven by the innovation of model architecture. For example, from statistical machine learning related to probability distribution functions to neural networks with linear transformations and nonlinear activations, the model complexity rises significantly to study more complicated problems. The proposal of residual learning remarkably deepens the neural networks, which makes them quite powerful in the initial image recognition problem to modern AI models~\cite{resnet2016}. The most outstanding breakthrough is probably the Transformer architecture~\cite{NIPS2017_attention}, which reshapes the realm of natural language processing and computer vision. Its capability of capturing the long-range dependencies between samples provides considerable power to generate new objects based on the sequential inputs. 
The integration of graph neural networks with transformer-based message passing, i.e., graph transformer, is playing an important role in various applications~\cite{dwivedi2026graphtransformer}.
Currently, there are several typical modular model architectures, for example, Multiple Layer Perceptron (MLP), Transformer and its derivatives, Mamba, Kolmogorov-Arnold Networks (KAN), Mixture-of-Experts (MoE), logistic regression (LR), and so on. How to adapt these architectures in applying AI for materials science is waiting to be explored, as well as asking for new architectures. Another crucial issue is how to assemble these architectures into a sophisticated model to handle materials science problems. This will rely on how deeply we understand the data and the domain knowledge.

The next component is Encoding, which demonstrates a numerical vector or tensor to represent an object. It is also called representation or embedding. Computers are good at fast calculations by processing binary digits, rather than any other realistic objects. To harness these computational powers, we first should transform them into a numerical format for processors. The embeddings of tokens in language models and the pixels of images in computer vision are prototypical examples. In fact, informative representations are crucial to AI development, especially in mining hidden insights and simplifying model architectures. Linear models should be the best as long as the representations carry sufficient messages.
In general, there are two methods of encoding~\cite{ouyang_graph_2025,atom2vec_pnas_2018}. One is manual design, in which the representation of an object is constructed by the concatenation of certain properties. This is straightforward and heavily employed in the literature. In early natural language processing, words in a document are usually represented by their occurrence frequency, i.e., the bag-of-words method. The other is dynamical learning. In modern LLMs, tokens from text-splitting are randomized, and then their representations are learned during training. During inference, their representations can be rather dynamic from the context, mediated by their input embeddings and model architecture. This strategy is more sophisticated and challenging than the former, requiring an extensive dataset and effective learning methods.
In materials science, at the current stage, the first way is dominant. Representations of materials are usually combinations of numerical properties of chemical elements and the associated compounds. There are some drawbacks to be aware of. Firstly, these components of the numerical representations are discrete in nature, and the completeness of the feature set can never be guaranteed. Secondly, there are hidden overlaps between these components, for example, the melting temperature and the boiling temperature. The material properties themselves actually require decent vectorial representations. 
There are meanwhile two strategies for ice-breaking. The first is a comprehensive data sampling to ensure continuous gradient change in the latent space. The second is embedding domain knowledge into the numerical representations, for example, physics-embedded feature design. The powerful text mining by LLMs grants great promises in this pathway. Learning dynamical representations of various fundamental materials shall be the ultimate goal.

The third component is Optimization. It provides the mathematical framework through which learning, training, and decision-making are defined and resolved. It indicates the learning process by feature matrix manipulation in the transformative latent spaces through minimizing a loss function. Most AI systems learn by optimizing an objective function—such as minimizing prediction error or maximizing expected reward—over model parameters or policies. The loss function is designed to reduce $|E_{\rm in}-E_{\rm out}|$ in equation~(\ref{eq1}) by guiding optimal solution search in the learning landscape. 
Defining an effective loss function is of paramount importance (see those of AlphaFold from ref.~\cite{Jumper2021HighlyAP}). It requires a clear understanding on the input data and the learning target, and a mathematical expression of the objective function. 
This perspective unifies diverse areas of AI, including statistical learning, deep learning, and reinforcement learning, all of which can be viewed as optimization problems under uncertainty and constraints. The choice of optimization formulation and algorithm strongly influences model performance, generalization, robustness, and scalability. 
Practically, it also determines the training efficiency and thus the cost.
In this sense, optimization is not merely a tool for AI, but a core component that shapes what intelligent systems can learn and how they behave.
Turning a handy problem into a learnable one is the key step in AI research and its applications.
In addition, optimization algorithms are also very important, of which gradient descent is the most well-established. Developing new algorithms will be helpful to boost the training efficiency. 
Currently, in materials science, this point is much less investigated. Quite a difference can be observed from the development of AlphaFold~\cite{Jumper2021HighlyAP}. To develop whatever specialized or foundational models, an effective loss function should be defined before learning.

%The last component is Inference. This is the final target of all research, i.e. to decode unknown from the known. In some sense, this is equivalent to Prediction or Generation. From the data itself, by sophisticated data mining, critical knowledge is predictable (see discussion above). In statistical machine learning, the predicted entity is either a specific class (classification) or a value in a numeric range (regression). Once the model is trained, the process of getting the label for the unknown data is called prediction. 
If data, architecture, encoding, and optimization represent the key components in model training, the last component, inference, is its moment of model usage---the critical juncture where a trained AI model is tasked with making predictions or generating on unseen data. In the context of materials science, inference is the tool that transforms a computational model from a passive repository of learned patterns into an active predictor of material properties or associated natural language description.
%It is the step that allows a researcher to query a model with the chemical formula of a novel amorphous alloy and receive a prediction of its glass forming ability, or to ask whether a proposed lithium-ion conductor is likely to be thermodynamically stable. 
Without efficient and reliable inference, even the most meticulously trained model remains nothing more than an academic artifact.
Inference becomes more prevalent in the era of LLMs. The autoregressive nature of LLMs generates the next token from the given context. The results are very dynamic. Therefore, the outcome from the model will strongly depend on the user input text, i.e., the prompt. How to design the prompt, called prompt engineering, is of critical importance to retrieve desired messages. It sounds like the model weights carry all the training information, and prompt engineering guides the model generation. The inference efficiency has been intensively discussed in science and industry, for which a lot of techniques have been proposed, such as the popular chain-of-thought~\cite{CoT_NIPS2022}. There are generally two ways for prompt engineering: (i) Frontend engineering, such as designing prompt templates, few-shot learning, retrieval-augmented generation (RAG), and low-rank adaptation. These methods do not change the model weights. (ii) Backend engineering, such as supervised fine-tuning, continued pre-training, and reinforcement learning from human feedback~\cite{deepseek_nature2025}. The model weights will be updated in the process.
These cases shall also apply to materials science regarding to either the foundation LLMs or domain-specific LLMs. The data format of interest can be quite different from language, which requires specific ways of prompt engineering. How to effectively extract the scientific knowledge from LLMs in scientific research is of great value in future investigations.

The four components and their combinations shall attract sufficient attention in AI research. The total power can be leveraged by strengthening either one. For example, if the entity representation (Encoding) is very accurate, the model architecture can be simplified to achieve similar performance. Improving the data quality and sampling will make the full process more lightweight. In future AI research, especially serving for materials science,  each node, each edge, each triangle, and the tetrahedron itself should be investigated comprehensively. This should provide clues to decompose difficult questions and design feasible pathways.

\subsection{Material network science}
In natural science, especially materials science, data is not always enough. The crucial points are the unknown nature of the potential space or landscape and the sampling efficiency from the current data. They are usually expensive experimental data that are hard to generate and reproduce. Meanwhile, these data have enriched the physical essence. They can be quite difficult to understand even for human brains with considerably trained neurons. This is mainly because of data scarcity and their discontinuous gradients in the dynamical latent space.
In addition, this data scarcity actually generates constraints on the usage of complex models (see discussion above). Currently, these data are often represented in a tabular format, which is learned by statistical machine learning. An incautious deep learning model is prone to cause overfitting, which makes the capability of generalization and prediction quite weak.

%%%%%%%%%%%%%%%%%%%%%%
\begin{figure}[t!]
\centering
\includegraphics[width=\textwidth]{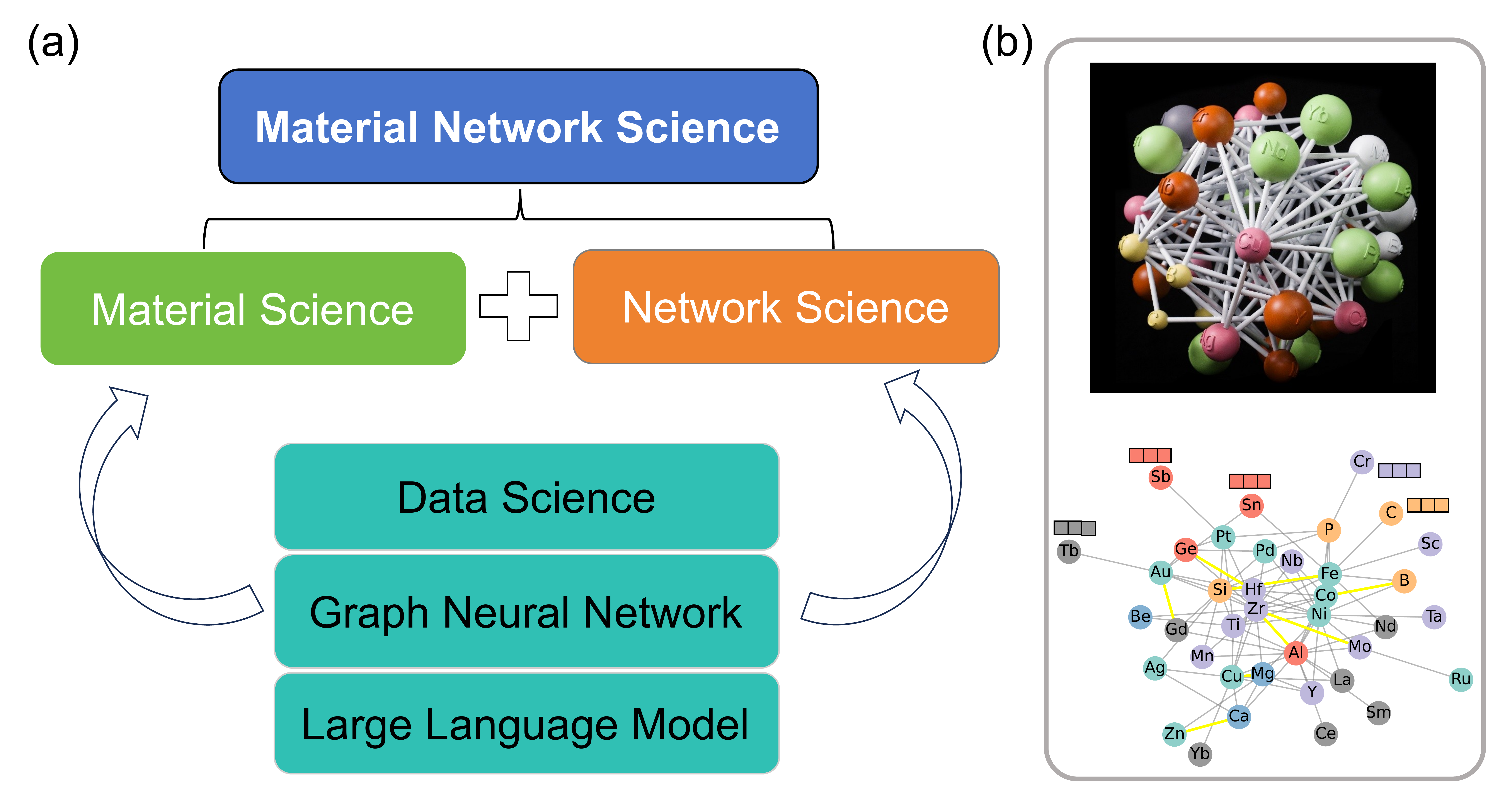}
\caption{
{\bf The protocol of Material network science to integrate AI and materials science}.
The left panel shows the general framework, and the right illustrates the material network of binary amorphous alloys from art design. Two layouts are shown, with the upper panel 3D printed and the lower 2D algorithmically generated.
}
\label{fig6}
\end{figure}
%%%%%%%%%%%%%%%%%%%%%%

To facilitate the integration of AI with materials science, we propose the material network science strategy to transform the scientific problem to sophisticated graph decoding one. The key idea is to combine materials science and network science, which incorporates data science, graph neural network, and LLM (see Fig.~\ref{fig6}a). 
The first step is data collection and cleaning, from which a three-dimensional network can be built. The nodes and edges are designed very flexibly, base on the research target. 
In Fig.~\ref{fig6}b, we show examples of materials networks designed for binary metallic glass-forming alloys. The upper panel is 3D-printed, while the lower is flatland algorithmically generated.
There are several benefits of designing this three-dimensional network instead of the tabular representation.
(1) The hidden relationships between nodes and other entities can be learned much easier learned based on the graph. The mathematical tools from graph theory apply to extracting relevant material information. Specific data science approaches can be developed for different domains.
(2) With the designed network, the substructures, like node, edge, triangle, etc., shall be encoded in various ways, for example, with the assistance of LLMs from text mining. The knowledge graph itself is an important tool in this research direction. For the same problem, graph infusion or integration may be promising in future research.
(3) Graph neural network has been attracting intensive research interests and are currently well-developed. From the graph structure, these algorithms are inherently applicable to learn desired knowledge. This relieves the overfitting risk to some extent. Furthermore, these models can be easily adapted to the latest techniques, such as their optimization with the Transformer.
There are many new things left to be explored.

In our recent work~\cite{zhang_constructing_2025}, we applied this strategy to study the intelligent design of amorphous alloys from the limited experimental data accumulated in the past 60 years. The material networks of binary and ternary systems are generated and 3D-printed, as shown in Fig.~\ref{fig6}b for the binary system. There are 38 nodes and 94 edges (binary systems). The ternary one is way more complex than expected. Even though there are only 47 nodes (chemical elements) and 352 triangles (ternary systems), the network is quite complicated. It carries the knowledge and efforts from the field over six decades, as well as some unknown knowledge. Graph data mining unveils the experimental alloy design strategies hidden in the data pattern and provides suggestions on new materials. It also identifies an "innovation trap" from the dynamical network analysis for some key obscured insights in the traditional research. This means this process can be accelerated by the network representation~\cite{shen_making_2025}. Moreover, the degree distribution results show the abnormal scale-free feature, which is sourced from the inherent physical constraints. Similar features are found in some other daily-life networks, typically, the fight networks.

In addition, with the networks learned handy, we encoded the chemical elements from text-processing of their Wikipedia contents~\cite{ouyang_graph_2025}. The properties of the chemical elements are embedded in the corresponding numerical representations. Even though not sufficient from the plain expression of the Wikipedia entities, this initiates the journey of elemental encoding with advanced multi-modal LLMs. This brings fascinating opportunities to adapt domain knowledge to feature engineering. By integrating graph neural networks, we trained a recommendation system to suggest material candidates on the graph structure. Tremendous efforts are ongoing to optimize the recommendation system. Akin to product suggestion in online shopping, this may be a promising protocol to enable smart material design.
These studies serve as a protocol example of applying material network science in addressing well-defined, resolvable material design challenges.

In principle, the concept of material network science can be extended to different fields and sub-tasks. The key in this line lies in the idea of dimensionality upgrade, which maps a low-dimensional hard problem to a learnable high-dimensional one. This, in some sense, aligns with the spirit of machine learning itself. How to design proper material networks in various domains is of practical importance. 
Extending to knowledge graphs and LLM domains is highly demanded.
We leave this exploration for broader creativity from the readers.

\section{Conclusion}
%Summarize your findings and contributions.
Materials science is an important subject in natural science, showing mysteries in many materials that are hard to decode by human intelligence. With the unprecedented success of AI techniques, especially the multi-modal LLMs, the way of scientific thinking and operation is reshaped. Not only the AI tools but also AI principles are of crucial importance to agitate next breakthrough in materials science. The classical material tetrahedron representing the Structure-Property-Processing-Performance-Characterization relationships has been providing guidelines in material research for a long time. By recalling its evolving trace and considering the emerging AI techniques, we propose new tetrahedral research paradigms for AI for materials science and the AI research itself that can be associated with materials science. One is the Matter-Data-Model-Potential-Agent diagram, and the other is the Data-Architecture-Encoding-Optimization-Inference relationship. Various research focus has been suggested, and a specific research strategy can be motivated. Along this line we introduce the material network science to help the effective integration of AI and materials science to foster breakthroughs.
% While AI is flourishing everywhere, including scientific research,
While AI is proliferating across scientific domains, we shall come up with well-defined, resolvable problems for materials science to harness the power of AI. This work shall motivate new systematic research of AI for science.

\section*{Acknowledgments}
%Thank collaborators, funding sources, etc.
This work is supported by the National Natural Science Foundation of China (Grant No. 52471178).

% Bibliography style
\bibliographystyle{naturemag_noURL}
\bibliography{refs}

\end{spacing}
\end{document}